\documentclass[%
 reprint,
superscriptaddress,
 amsmath,amssymb,
 aps,prl,
8pt]{revtex4-1}

\usepackage{graphicx}
\usepackage{dcolumn}
\usepackage{bm}
\usepackage{physics}
\usepackage{float}
\usepackage{rotating}
\usepackage{subfig}
\usepackage{color}
\usepackage{array}
\usepackage{makecell}
\usepackage{ragged2e}
\usepackage[font=footnotesize,
   justification=centerlast, singlelinecheck=off, textformat = simple, format = plain]{caption}

\newcommand{\BPCB}{(C$_5$H$_{12}$N)$_2$CuBr$_4$}

\begin{document}

\preprint{APS/123-QED}

\title{Magnetic field induced bound states in spin-$\frac{1}{2}$ ladders}

\author{Mithilesh Nayak}
\email{mithilesh.nayak@epfl.ch}
\affiliation{Institute of Physics, Ecole Polytechnique F\'ed\'erale de Lausanne (EPFL), CH-1015 Lausanne, Switzerland}

\author{Dominic Blosser}
\email{dblosser@phys.ethz.ch}
\affiliation{Laboratory  for  Solid  State  Physics,  ETH  Z\"urich, CH-8093  Z\"urich,  Switzerland}


\author{Andrey Zheludev}
\affiliation{Laboratory  for  Solid  State  Physics,  ETH  Z\"urich, CH-8093  Z\"urich,  Switzerland}

\author{Fr\'ed\'eric Mila}
\affiliation{Institute of Physics, Ecole Polytechnique F\'ed\'erale de Lausanne (EPFL), CH-1015 Lausanne, Switzerland}%

\date{\today}

\begin{abstract}
Motivated by the intriguing mode splittings in a magnetic field recently observed with inelastic neutron scattering in the spin ladder compound (C$_5$H$_{12}$N)$_2$CuBr$_4$ (BPCB), we investigate the nature of the spin ladder excitations using DMRG and analytical arguments. Starting from the fully frustrated ladder, for which we derive the low-energy spectrum, we show that bound states are generically present close to $k=0$ in the dynamical structure factor of spin ladders above $H_{c 1}$, and that they are characterized by a field-independent binding energy and an intensity that grows with $H-H_{c 1}$. These predictions are shown to explain quantitatively the split modes observed in BPCB.
\end{abstract}

\maketitle

Intermediate between chains and 2D lattices, spin ladders have played a central role in the investigation of quantum effects in magnetism \cite{Dagotto618}. Being effectively 1D, they can be studied by state-of-the-art methods of quantum 1D physics such as bosonization \cite{Schulz1986, StrongMillis1992, Giamarchi2003} and density matrix renormalization group (DMRG) \cite{WhiteDMRG1992,Schollwoeck2011}.
On the experimental side, numerous spin ladder compounds have been discovered and extensively studied \cite{ Kato1998, Cavadini2001,Watson_Kotov2001, LorenzHeyer2008, AnfusoGarst2008, ThielemannRueggKiefer2009}. 
Accordingly, their physics is very well understood. In zero field, the spectrum of a ladder is gapped regardless of the ratio of leg to rung coupling \cite{Giamarchi2003}, and when they are of the same order, the ground state can be interpreted as some kind of resonating valence bond state, implementing an idea put forward by Anderson in his seminal paper on the theory of high T$_c$ superconductors \cite{Anderson1987}. The spin gap can be closed by a magnetic field, leading to a Tomonaga-Luttinger liquid phase with gapless excitations and incommensurate correlations \cite{Dagotto_Riera_Swanson_1993, Chitra_Giamarchi_1997, Giamarchi_Tselvik_1999,Giamarchi_Ruegg_review2008, Ward2012,Ruegg2008,Klanjsec2008NMROrdering,Sologubenko2009}. All these properties have been observed in many systems. For instance, inelastic neutron scattering (INS) has been used to study the fate of the low-lying excitations across the quantum phase transition at which the gap closes \cite{Savici2009, ThielemannRuegg2009, Schmidiger2013, Povarov2015}, and a continuum of excitations has been observed in the gapless phase, in perfect agreement with theoretical expectations. It has even been possible to check the universal finite-temperature scaling of the transverse local dynamical structure factor at the gapped--gapless quantum critical point \cite{Blosser2018}.

\begin{figure}
\centering
\includegraphics[width = 7.0cm , height = 7.0cm , keepaspectratio]{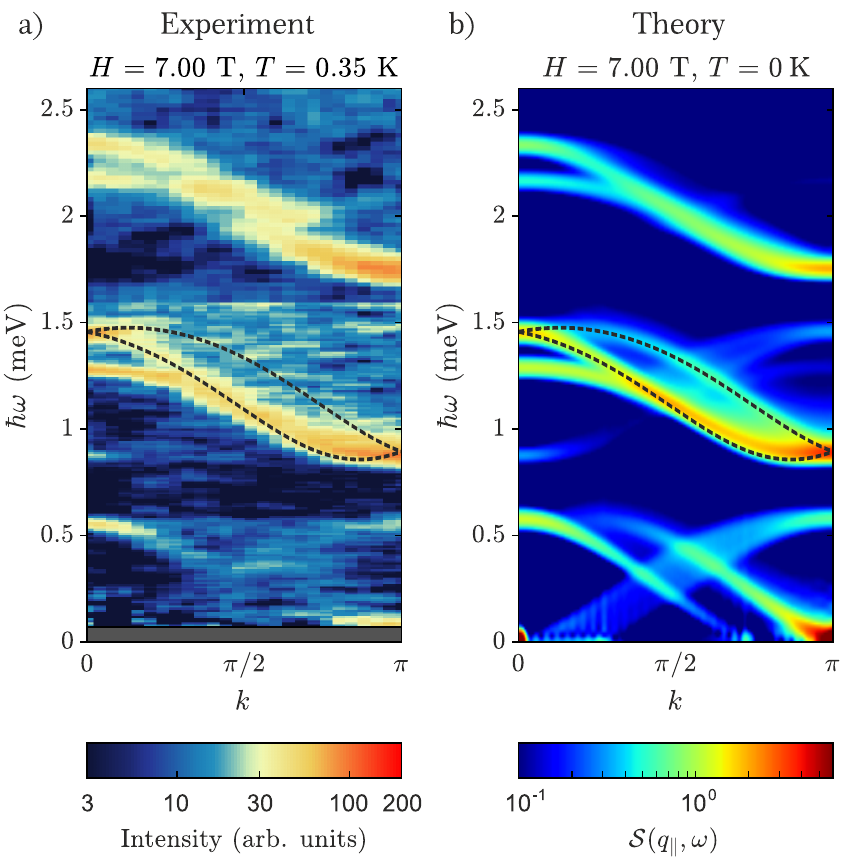}
\caption{
(a) INS spectrum measured in BPCB in the slightly magnetized regime. Three strongly Zeeman-split triplet branches are clearly visible. Near $k=0$ a distinct splitting of the middle and upper triplet band is observed. These data were previously published in Ref.~\cite{Blosser2018}.
(b) DMRG simulation of the dynamic structure factor calculated at the same magnetic field. The black dashed lines in the DSF plots of both  panels mark the boundaries of the continuum associated to the dispersion of $t^0$ rung triplets as predicted by the mapping on the $t-J$ model\cite{Bouillot2011}.}
\label{fig1}
\end{figure}

\begin{figure}
\centering
\includegraphics[width = 8.5cm , height = 8.5cm , keepaspectratio]{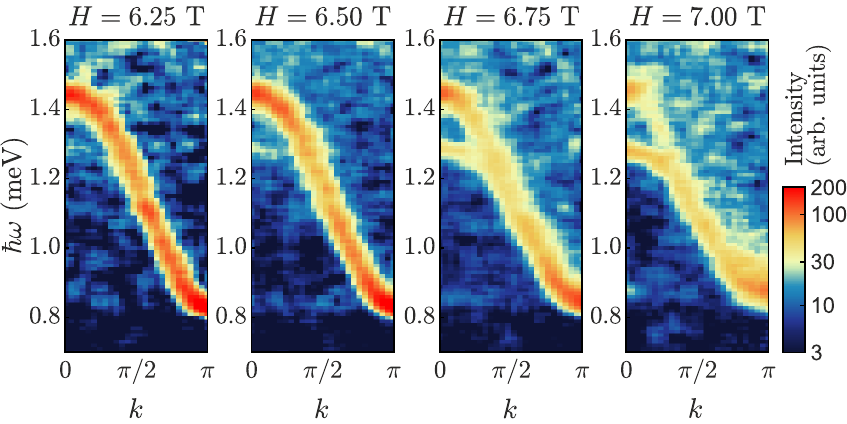}
\caption{Evolution of the INS spectrum measured in BPCB at 0.35~K at different magnetic fields near the critical field of $H=6.66(6)$~T. In the plotted range of energy transfer only the middle triplet branch of excitations is visible. Upon increasing the magnetic field beyond $H_{c1}$, a distinct splitting is observed near the band maximum. These data were previously published in Ref. \cite{Blosser2018}.}
\label{fig2}
\end{figure}

Some puzzles remain however. One of them has to do with the higher energy excitations in the gapless phase. In the presence of a magnetic field, triplet excitations are split. The lowest branch crosses the singlet ground state and gives rise to the low-energy continuum, while the other two branches lie at a finite energy. The dynamical spin structure factor of these branches turns out to have a rather complex structure however, as revealed by neutron scattering on (C$_5$H$_{12}$N)$_2$CuBr$_4$ (BPCB) and time-dependent DMRG \cite{ Daley_2004, White_Feiguin2004, Barthel_Schollwoeck_White2009} on a ladder with strong rungs. The main features, that will be discussed in great details below, are summarized in Fig.~\ref{fig1} and \ref{fig2}. Let us concentrate on the intermediate branch. It consists of three main features, and a weak continuum barely visible on that scale. The two upper features (black dashed lines in Fig.\ref{fig1}), and the associated continuum, have been nicely explained by Bouillot \emph{et al.} \cite{Bouillot2011} in terms of an effective $t-J$ model where the ground state is described by an effective spin chain with a finite magnetization, 
and the $S_z=0$ triplet excitation is represented by a mobile hole.

By contrast, the lowest feature of the intermediate branch, the split mode around $k=0$ below the continuum bounded by the black dashed lines in Fig.\ref{fig1}, has not been discussed. This splitting, which is also present in the upper branch, clearly calls for an explanation.

In this Letter, we provide strong evidence that this split mode is a bound state of the excited triplet with one of the triplets induced in the ground state by the magnetic field in the gapless phase. This identification relies on three results: i) The exact solution of the low-lying spectrum of the fully frustrated ladder with cross couplings equal to leg couplings in terms of triplets and of bound states of pairs of triplets; ii) The smooth evolution of the lowest bound state upon reducing frustration, as revealed by extensive DMRG simulations; iii) The prediction that the intensity of this mode grows with the distance to the critical field while its binding energy is essentially independent of the field, a prediction in perfect agreement with the neutron scattering results on BPCB in the gapless phase. 
Note that bound states have been discussed before in zero magnetic field \cite{Damle_Sachdev1998, Sushkov_Kotov1998, Sushkov_Kotov1999, Trebst2000} as multi-triplet excitations from the singlet ground state. To the best of our knowledge, the possibility of observing them above the first critical field with inelastic neutron scattering as split features of the main triplet bands with significant spectral weight had not been anticipated however.

\begin{table}
\centering
\caption{Action of the rung operators on a single rung.}
\begin{tabular}{c|c|c|c|c}

 & $S^{z}_{0}$ & $S^{z}_{\pi}$ & $S^{x}_{0}$ & $ S^{x}_{\pi}$\\
\hline
$ |s\rangle$  &$ 0$   &$ |t^0\rangle$ &$0$ &$ -\sqrt{2}(|t^{+}\rangle-|t^{-}\rangle)$ \\

 $ |t^{+}\rangle$ &$ |t^{+}\rangle$  &$ 0$  &$ \sqrt{2}|t^{0}\rangle$ &$ -\sqrt{2}|s\rangle$\\

 $ |t^0\rangle$ &$ 0$ &$ |s\rangle$ &$ \sqrt{2}(|t^{+}\rangle+|t^{-}\rangle)$ &$ 0$\\

 $ |t^{-}\rangle$ &-$ |t^{-}\rangle$ &$ 0$ &$ \sqrt{2}|t^0\rangle$ &$ \sqrt{2}|s\rangle$\\
 
\end{tabular}
\label{Table1}
\end{table}

Our experimental results were obtained on the compound \BPCB{} (BPCB), a particularly well-studied $S=1/2$ strong rung spin ladder material \cite{Patyal1990, Ruegg2008, Klanjsec2008NMROrdering, Savici2009,  ThielemannRuegg2009, Blosser2018}. The ladders are formed by magnetic Cu$^{2+}$ cations and linking Br$^-$ anions \cite{Patyal1990}. The rung and leg couplings respectively are given by $J_\perp= 12.67(6)$ K and $J_\parallel =3.54(3)$ K and the critical field is $\mu_0 H_\mathrm{c1}=6.66(6)$~T \cite{Blosser2018}. All the inelastic neutron scattering data \cite{DataISIS_BPCB} were previously published in Ref.~\cite{Blosser2018}, and experimental details can be found therein. The mentioned study exclusively focused on the universal low energy spin dynamics. The subject of the present paper on the other hand is the high energy triplet excitations. 

All theoretical results reported in this paper have been obtained on the model 
\begin{eqnarray*}
\cal{H} & = J_{\parallel}\sum_{ i,j = 1,2} \vec S_{i,j}\cdot \vec S_{i+1,j} + J_{\perp}\sum_{i} \vec S_{i,1} \cdot \vec S_{i,2} \nonumber \\
+  J_{\cross} & \sum_{i} (\vec S_{i,1}\cdot \vec S_{ i+1,2}+\vec S_{i,2}\cdot \vec S_{i+1,1} )- h\sum_{ i,j=1,2} S^{z}_{i,j}
\label{Hamiltonian}
\end{eqnarray*}
where $J_{\perp}$ is the rung coupling, $J_{\parallel}$ is the leg coupling, and $J_{\cross}$ is a cross coupling (not present in BPCB) that frustrates the leg coupling. The component of the vector operators $\vec S_{i,j}$ are spin-1/2 operators. The first index keeps track of the rung, the second one of the leg. The $g$ factor and the Bohr magneton $\mu_B$ have been included in the magnetic field $h$. The numerical results have been obtained using the time-dependent DMRG method pioneered in this context by Bouillot \emph{et al.} \cite{Bouillot2011}, and we have benchmarked our code by reproducing the results of Bouillot \emph{et al.} in the unfrustrated case ($J_{\cross}=0$) \cite{Bouillot2011}. The dynamical structure factor (DSF) can be defined in the Lehmann representation by
\begin{equation*}
S^{\alpha \alpha}_{k_{\perp}}(k,\omega) = \frac{2\pi}{N_r}\sum_{\eta} {|\bra{\eta}S^{\alpha}_{k_{\perp}}(k)\ket{\mathrm{GS}}|}^2 \delta(\omega+E_{\text{\tiny GS}}-E_{\eta})
\end{equation*}
where $N_r$ is the number of rungs, $\ket{GS}$ is the ground state and the sum over $\eta$ runs over the excited states. The rung operators (in position basis) are defined as $S^{\alpha}_{i, k_{\perp} = 0, \pi} = S^{\alpha}_{ i,1}\pm S^{\alpha}_{i,2}$. The action of the various components on the eigenstate of a rung dimer are summarized in Table \ref{Table1}. 

In Fig.~\ref{fig1}b), a false color plot of the sum of the longitudinal and transverse symmetric and antisymmetric DSF components is shown. The field has been adjusted to the experimental value of Fig.~\ref{fig1}a). These results agree qualitatively with those of Bouillot \emph{et al.}, obtained at a slightly larger field.
The inelastic neutron scattering cross-section contains the same DSF components with $k$-dependent weights  \cite{Schmidiger2013,Blosser2018}. In the scattering geometry of the present experiment, these weights are near unity. Further, the experimental and numerical resolution are quite comparable. Indeed, we find good qualitative agreement between the experiment and the numerical simulation.

\begin{figure*}
\centering
\includegraphics[width = 17.5cm , height = 17.5cm , keepaspectratio]{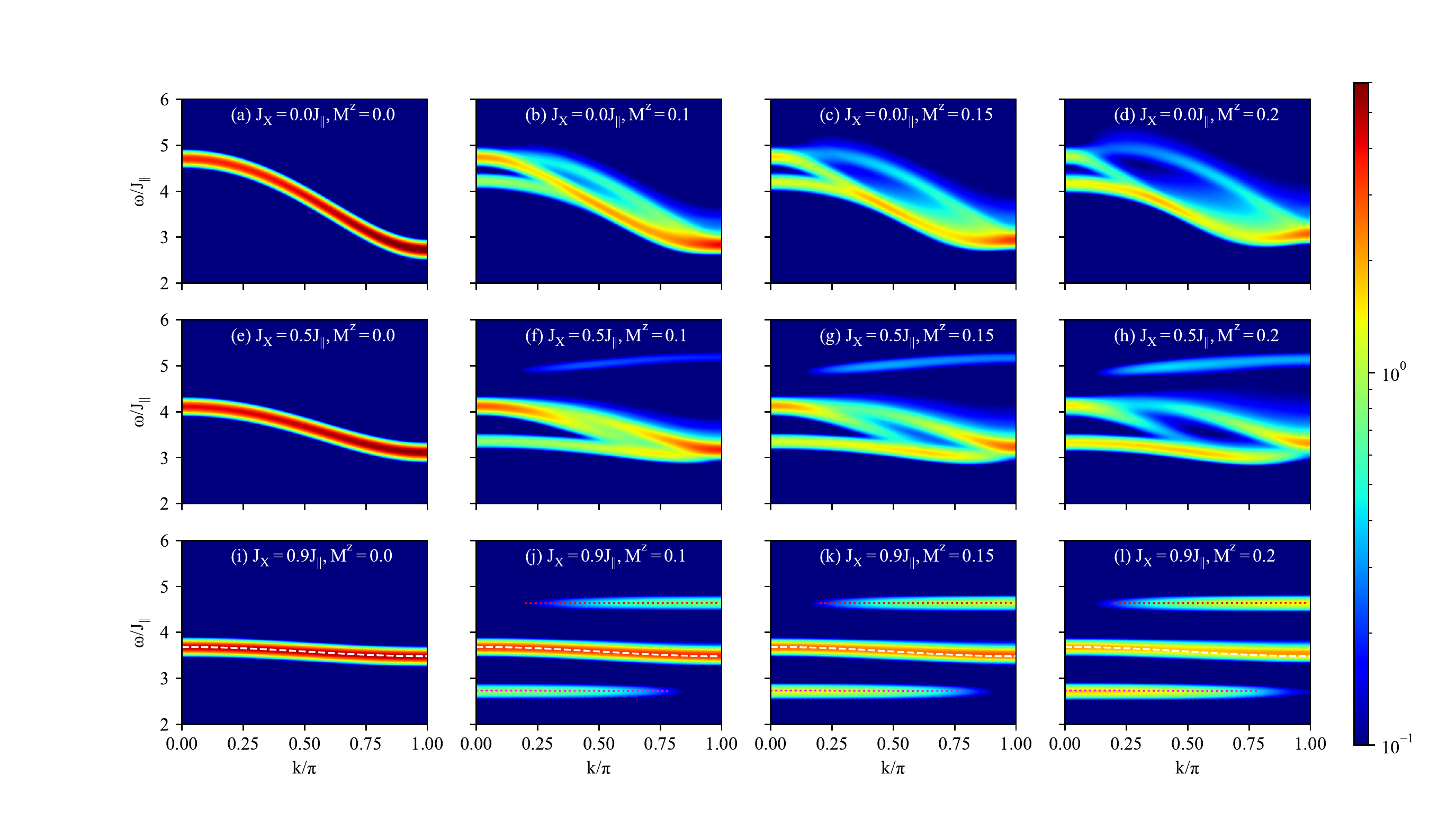}
\caption{Evolution of the longitudinal antisymmetric DSF $S^{zz}_{\pi}(k,\omega)$ with magnetization and frustration as obtained with DMRG (color plots) and perturbation theory (dashed lines, lower panels). At strong frustration, the three branches correspond to a single-triplet branch (middle) and two bound states with total spin 2 (upper mode) and total spin 1 (lower mode). Upon reducing the frustration, the upper bound state progressively loses its intensity, while the lower bound state evolves smoothly into the split mode at $k=0$ of the unfrustrated case, leading to the interpretation of this mode as a spin-1 bound state.}
\label{fig3}
\end{figure*}

Let us start by discussing the spectrum of the fully frustrated ladder ($J_{\cross}=J_{\parallel}\equiv J$) in the absence of a field \cite{Gelfand1991, Indrani1993, Xian1995, Honecker2000, Honecker_Wessel2016, Honecker_Mila_Normand2016}. The Hamiltonian can be rewritten entirely in terms of the sum of the spins on each rung so that the total spin of each rung is a conserved quantity. Accordingly, there are $2^{N_r}$ decoupled sectors of the Hilbert space, corresponding to the possible values of the spin (0 or 1) of each rung. For $J_\perp/J>1.401$, the ground state is a product of singlets. 
The lowest energy excitation consists of creating a triplet on a rung. It is immobile because both neighbors are singlets, its energy is equal to $J_\perp$, and it is $3\times N_r$-fold degenerate, the factor 3 coming from the 3 possible values of $S_z$, -1, 0, or 1. The next excitations correspond to the creation of two triplets. If the triplets are separated by at least one singlet, they are immobile. The energy of such an excitation is equal to $2J_\perp$, and it is $9 \times N_r(N_r-3)/2$-fold degenerate. By contrast, if the triplets are nearest neighbors, the spectrum is that of two $S=1$ spins coupled by $J$. The pair is immobile because it is surrounded by singlets. This will result in a singlet, a triplet, and a quintuplet bound state, with energies $2 J_\perp-2J$, $2 J_\perp-J$, and $2 J_\perp+J$, respectively. Each of these bound states is $N_r$ fold degenerate. The wave functions of these states are given in the Supplementary Material (see Ref. \cite{SM}). Higher energy states will not enter the discussion of the DSF, so we do not quote them.

In a magnetic field, and for $J_\perp$ large enough, the fully frustrated ladder has a direct transition from the product of singlets to a $1/2$ plateau state with triplets on every other bond at $h=J_\perp$ \cite{Totsuka1998, Mila1998, FouetMila2006}, followed by another direct transition to the fully saturated state at $h=J_\perp+2J$. So to discuss the DSF at small magnetization, we need to leave the fully frustrated limit and restore a continuous magnetization curve by introducing a small perturbation defined by $\delta J = J_{\parallel} - J_{\cross}$ such that, for $\delta J=0$, the ladder is fully frustrated, while for $\delta J=J_{\parallel}$, we get the regular ladder relevant for BPCB. In zero field, the degeneracy of the first excited states is lifted to first order, and the excitations acquire a dispersion given by (see \cite{SM})
$$
\omega_1(k)= J_\perp+ \delta J \cos k.
$$
For the two triplet excitations, the excitations corresponding to triplets far apart give rise to a two-magnon continuum built out of the dispersion $\omega_1(k)$. By contrast, the bound states acquire a dispersion, but since the basic process induced by $\delta J$ is to let a triplet hop, moving a pair is a second-order process, and the effective hopping is of the order $(\delta J)^2/J$ \cite{SM}. 
In a magnetic field, the first level to cross the singlet ground state is the one triplet level sitting at the bottom of the band. The only competitors are the two triplet states with $S_z=2$. Since the spin-2 bound state is higher than the state with two triplets far apart, the next state to cross will be a two-magnon state. This state will cross slightly later than the one triplet state because its energy is more than twice the ground state of the single triplet states - these triplets behave roughly speaking as spinless fermions with large nearest neighbour repulsion. Upon increasing the magnetization, the same argument will carry over with many-magnon states. The ground-state  magnetization increases smoothly \cite{SM}, and the ground state wave function is essentially a linear combination of states with the appropriate number of isolated triplets.

On the basis of these simple arguments, we are now in the position to predict the form of the DSF close to the fully frustrated limit. In the fully frustrated case, and in the limit $J=0$, the only excitation that has a non-zero $k_\perp=\pi$ matrix elements with a state with a few isolated triplets and that has energy $J_\perp$ corresponds to a transition from the singlet to the $t^0$ triplet ($S_z=0$) induced by $S_\pi^z$. This band will be split into three bands upon introducing $J$: a band of isolated triplets at energy $J_\perp$, and two bands of bound states corresponding to the states that contain the configuration $|t^{+}t^0\rangle$ or $|t^0t^{+}\rangle$, namely $|S=1,S^z=1\rangle =  (|t^{+}t^0\rangle-|t^0t^{+}\rangle)/\sqrt{2}$ and  $|S=2,S^z=1\rangle =  (|t^{+}t^0\rangle+|t^0t^{+}\rangle)/\sqrt{2}$,
of energy $J_\perp-J$ and $J_\perp+J$ respectively. Upon introducing the perturbation $\delta J$, the isolated triplet acquires a dispersion of amplitude $\delta J$, while the bound states acquire a dispersion of order $(\delta J)^2/J$, with an intensity proportional to $1+\cos k$ (resp. $1-\cos k$) for the lower (resp. upper) mode and a momentum shift by $\pi$ \cite{SM}. Finally, intensity of the isolated triplet is expected to be much larger at low magnetization because the probability to excite a singlet not adjacent to a triplet is almost equal to 1 (it is 1 in the limit of vanishing magnetization), while the bound states are expected to have an intensity proportional to the number of triplets in the ground state, i.e. to the magnetization. 

\begin{figure}
\centering
\includegraphics[width = 8.5cm , height = 8.5cm , keepaspectratio]{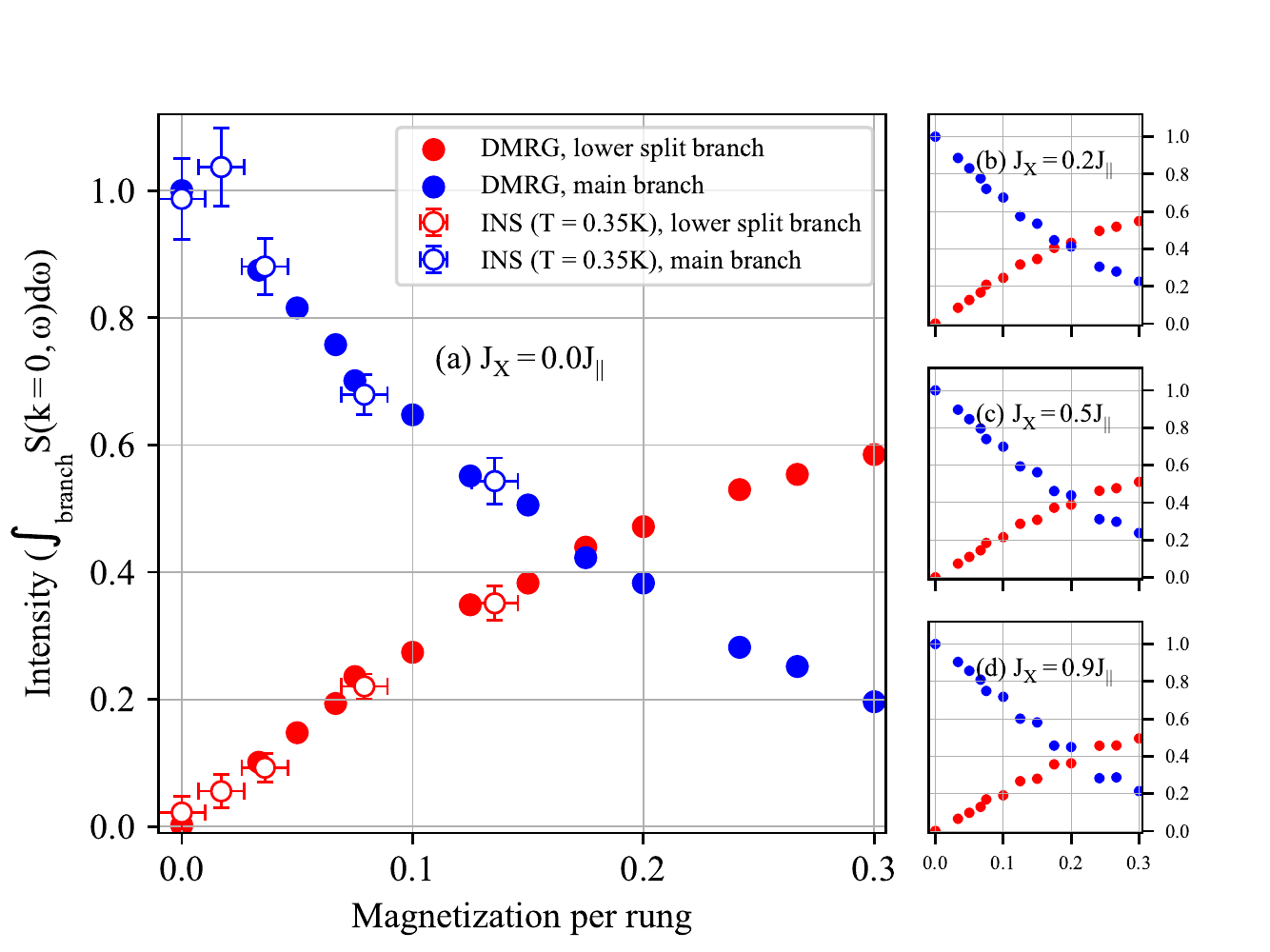}
\caption{Main panel: Intensity of the two main branches at $k=0$ as a function of magnetization from DMRG (solid symbols) and experiments (open symbols). The agreement between theory and experiment is quantitative, and the main effects are very clear: the intensity of the split mode increases with magnetization, as it should for a bound state, while that of the single triplet branch decreases. Right panels: DMRG results for frustrated ladders. The results are qualitatively the same.}
\label{fig4}
\end{figure}

These predictions are fully supported by DMRG simulations (see Fig.~\ref{fig3}). The spectrum of a ladder close to full frustration ($\delta J=0.1 J$) is shown in the bottom panels as a function of magnetization. As expected, a single branch with small but visible dispersion is present at zero magnetization. Upon increasing the magnetization, two additional branches appear with essentially no dispersion, and with an intensity that increases with the magnetization.

To make connection with the standard ladder relevant to BPCB, we have performed extensive DMRG simulations for several values of $J_{\cross}$. The spectrum acquires a complicated structure, but the evolution from the fully frustrated limit is transparent: i) Two copies of the single triplet band appear, shifted by a vector that increases with the magnetization, as predicted by the $t-J$ model analogy \cite{Bouillot2011}; ii) The upper bound state loses its intensity and is no longer visible in the standard ladder; iii) The lower bound is unaffected close to $k=0$ while for larger $k$ it gets mixed with the dispersive single triplet excitation.

These results strongly suggest that the split mode is the $S=1, S_z=1$ bound state of the fully frustrated case that survives as a well defined excitation close to $k=0$. To lend further support to this interpretation, let us look at the position and intensity of this branch. In the limit of full frustration, the position of this bound state is at $J$ below the single triplet branch independent of the magnetic field, while its intensity grows linearly with magnetization. 

These properties are shared by the split mode at $k=0$ of both the DMRG results of the standard ladder and the INS results on BPCB (see Fig. \ref{fig2}) with a remarkable degree of accuracy. Indeed, the energy of the split mode does not change in any significant way from its value when it first appears. By contrast, its intensity increases quite fast with magnetization, as shown in Fig. \ref{fig4}, while at the same time the intensity of the single triplet mode decreases. The DMRG simulations show that this is essentially independent of frustration (see Figs. \ref{fig4} b-d), supporting the continuity between the fully frustrated and the standard limits, hence the interpretation as a bound state. The comparison between the standard ladder and BPCB is shown in the main panel. To make this plot,  the experimental branch intensities have been extracted from narrow slices $k/2\pi\in \left[ -0.05,0.05\right]$ of the neutron scattering data obtained at $\mu_0 H=6.00$, 6.25, 6.50, 6.75 and 7.00~T. The corresponding magnetization was directly measured at the same fields using a Faraday balance magnetometer (see \cite{SM}). After multiplying the arbitrarily normalized neutron scattering data with a single overall prefactor, these points have been included in Fig.~\ref{fig4} as open symbols. The DMRG intensities have been obtained along similar lines \cite{SM}.

The same kind of analysis can be made for the upper branch of excitation. This branch corresponds to the excitation of a rung singlet to the $S^z=-1$ triplet under the action of $S^x_\pi$. In the fully frustrated case, it gives rise to a main branch at $2J_\perp$, and to three bound state branches corresponding to the singlet, triplet, and quintuplet bound states of energy $2 J_\perp-2J$, $2 J_\perp-J$, and $2 J_\perp+J$ since all of them have a component with $|t^{+}t^{-}\rangle$. Upon reducing frustration, the only boundstate that remains visible is the triplet with energy $2 J_\perp-J$, giving rise to a split mode very similar to that of the intermediate branch \cite{SM}.

In conclusion, we have solved one of the main remaining open questions in the physics of spin-1/2 ladders, the origin of the split modes observed in inelastic neutron scattering. Starting from the fully frustrated ladder, in which single particle and bound state excitations are very transparent, we have shown by continuity using DMRG simulations that the split modes originate from spin-1 bound states. It would be interesting to see if this conclusion can also be reached as a consequence of the nearest-neighbor attractive interaction between a hole and the up triplet in the context of the $t-J$ description of the excitations \cite{SM}. This goes beyond the scope of the present paper however. 

\paragraph{Acknowledgements:}
We would like to thank Vivek K. Bhartiya and D. J. Voneshen for their help with the experiment. 
This work is partially supported by the Swiss National Science Foundation under Division II. The numerical calculations have been performed using the facilities of Scientific IT and Application Support Center of EPFL.

\bibliography{Draft_final4.bbl}

\end{document}


\preprint{APS/123-QED}

\title{Supplemental Material for:\\
"Magnetic field induced bound states in spin-$\frac{1}{2}$ ladders"}

\author{Mithilesh Nayak}
\email{mithilesh.nayak@epfl.ch}
\affiliation{Institute of Physics, Ecole Polytechnique F\'ed\'erale de Lausanne (EPFL), CH-1015 Lausanne, Switzerland}

\author{Dominic Blosser}
\email{dblosser@phys.ethz.ch}
\affiliation{Laboratory  for  Solid  State  Physics,  ETH  Z\"urich, CH-8093  Z\"urich,  Switzerland}


\author{Andrey Zheludev}
\affiliation{Laboratory  for  Solid  State  Physics,  ETH  Z\"urich, CH-8093  Z\"urich,  Switzerland}

\author{Fr\'ed\'eric Mila}
\affiliation{Institute of Physics, Ecole Polytechnique F\'ed\'erale de Lausanne (EPFL), CH-1015 Lausanne, Switzerland}%

\date{\today}

\begin{abstract}
In this Supplemental Material, we give details about the way the magnetization has been measured and calculated, we explain how the intensity has been extracted from the DMRG data for the Dynamical Structure Factor, we derive the dispersion of the modes of the intermediate branch away from the fully frustrated limit up to second order in perturbation theory, and we show that the splitting of the higher energy branch can also be attributed to a bound state.
\end{abstract}

\maketitle

\section{Magnetization}

\subsection{Measurements}
\begin{figure}
\centering
\includegraphics[width = 10.5cm , height =10.5cm , keepaspectratio]{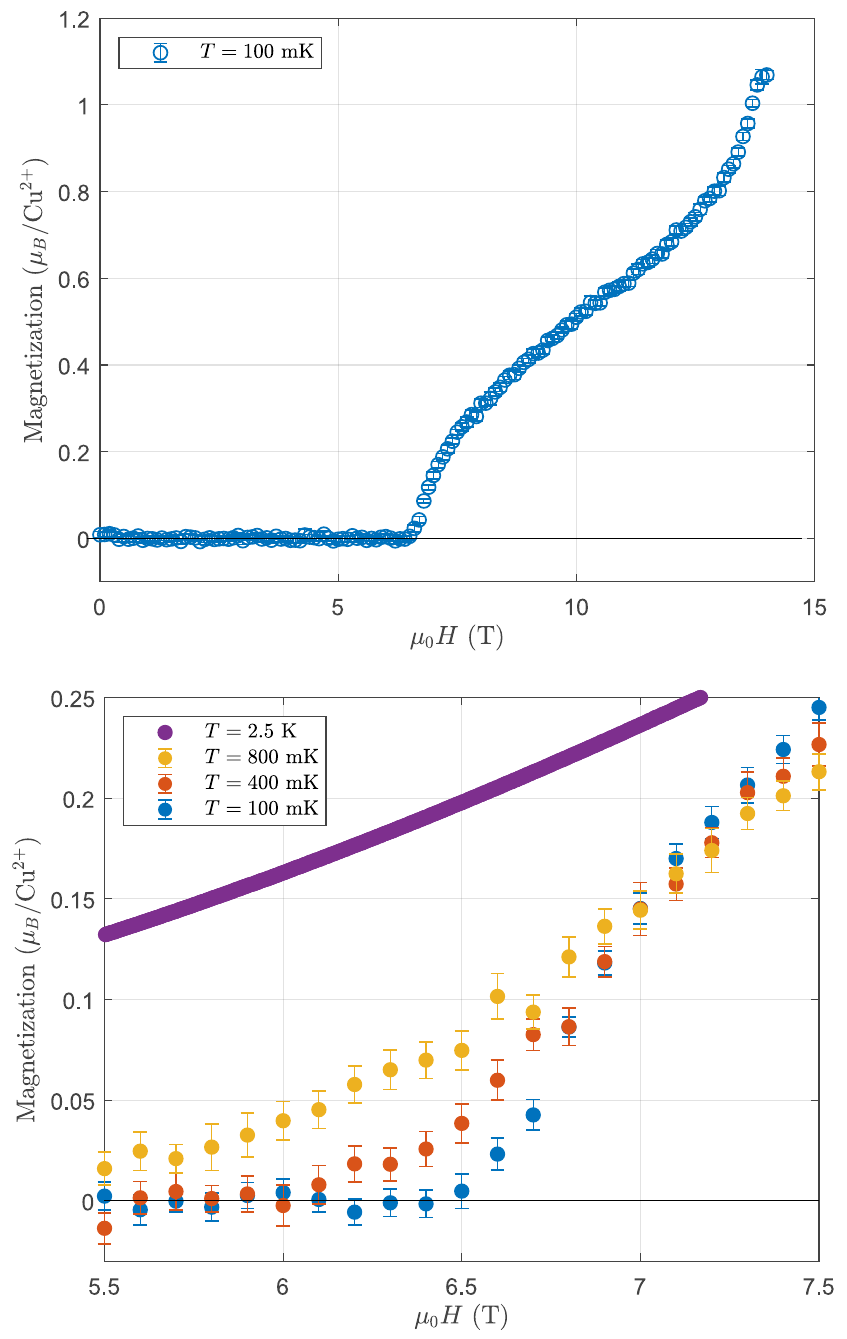}
\caption{Magnetization measured on a BPCB single crystal for $H\parallel b$. (a) Overview of magnetization at 100~mK. (b) Magnetization near $H_{c1}$ at various temperatures.}
\label{Supplfig2}
\end{figure}

Magnetization of a small ($\sim0.7$~mg) BPCB single crystal was measured using a custom built capacitive Faraday balance magnetometer. The sample was oriented with the  crystallographic $b$ direction parallel to the applied magnetic field, exactly as in the neutron scattering experiments \cite{Blosser2018}. Data were collected at $T=100$, $400$ and $800$~mK, as well as at several temperatures above 2~K. The absolute value of magnetization was obtained by matching the data collected above 2~K to measurements performed with a commercial (Quantum Design) vibrating sample magnetometer. An overview of magnetization measured at 100~mK up to $H=14$~T is shown in Fig.~\ref{Supplfig2}a). Fig.~\ref{Supplfig2}b) shows magnetization measured at various temperatures near the critical field $H_{c1}=6.66(6)$~T.
The magnetization values corresponding to the magnetic fields and temperatures at which the neutron scattering data were obtained, we read off these data. We note that the neutron scattering data was obtained at 350~mK, whilst magnetization was measured at 400~mK. Considering the sizeable error bars, however, this discrepancy is negligible. For comparison to the calculated values of magnetization, the experimental data is normalized by the magnetization at saturation $m_\mathrm{sat}$.

\subsection{DMRG results}

\begin{figure}
\centering
\includegraphics[width = 7.25cm , height =7.25cm , keepaspectratio]{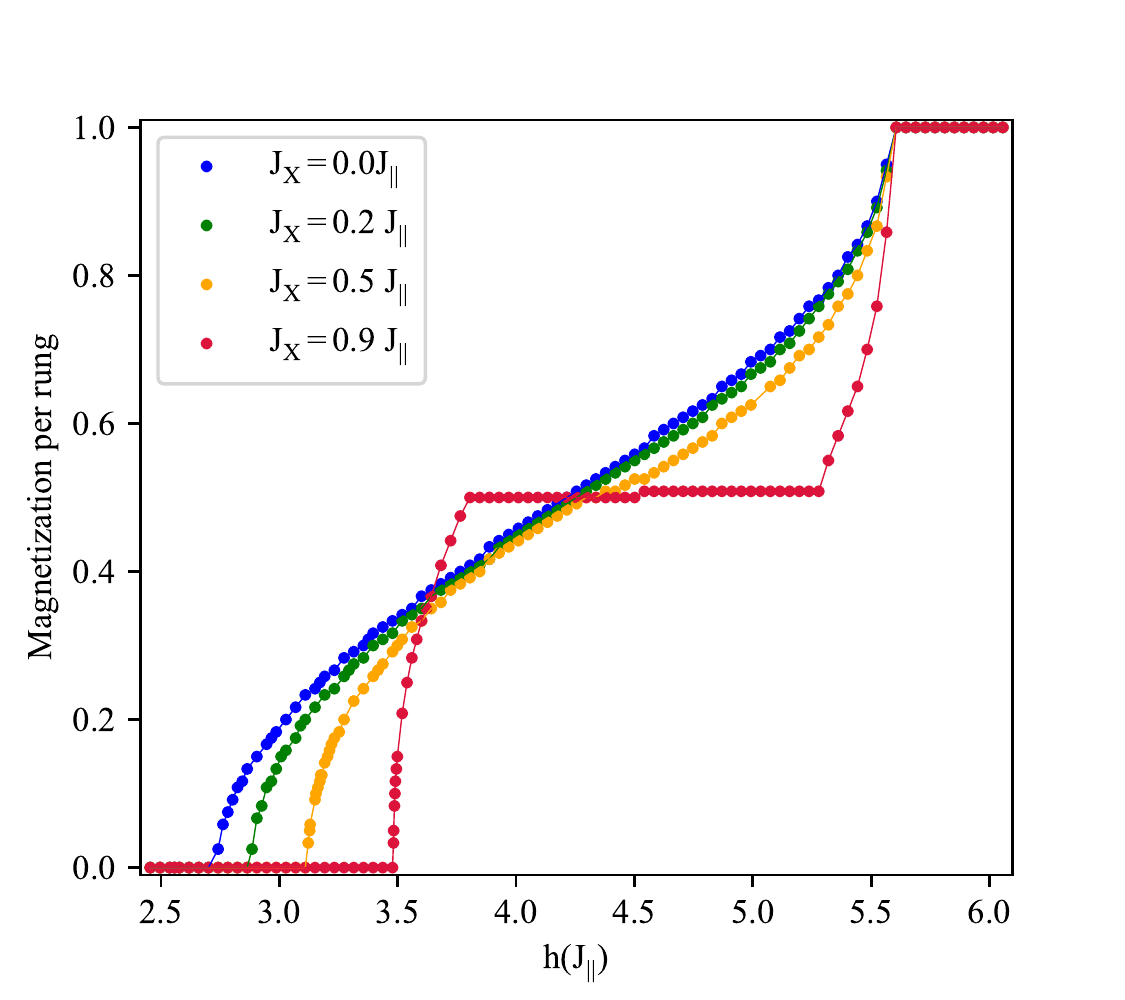}
\caption{Rung Magnetization vs. Magnetic Field. We find that the first critical magnetic field ($h_{c1}$) increasing with respect to increase in frustration. The magnetization plateau is revealed at a higher value of frustration as seen here. }
\label{Supplfig3}
\end{figure}

To discuss the evolution of the dynamical structure factor between the unfrustrated case and the fully frustrated case, it is more appropriate to work at fixed magnetization. However the DMRG simulations are performed at fixed field. So we have calculated the magnetization curve $M^z$ vs. $h(J_{\parallel})$ for several frustrated ladders in order to know how to fix the field to work at a given magnetization. The results are plotted in Fig. \ref{Supplfig3}. These DMRG simulations have been done on 120 rungs keeping 100 states. We find that the magnetization increases smoothly between two critical points $h_{c1}$ and $h_{c2}$ when the frustration is small. The ground state of the model below $h_{c1}$ is essentially a product of rung singlets that undergoes a transition to a gapless phase in which the groundstate is composed of singlets and $t^{+}$-triplets \cite{Gelfand1991, Xian1995, Honecker_Wessel2016}. Beyond $h_{c2}$, it undergoes another transition into a gapped phase in which the ground state is composed of $t^{+}$-triplets with rung magnetization 1. For larger frustration, a 1/2 magnetization plateau appears, leading to 4 critical points. The ground state in the magnetization plateau essentially consists of alternating rung singlets and $t^{+}$-triplets.

Close to the fully frustrated case, the first critical field $h_{c1}$ can be estimated from the dispersion of one triplet \cite{Reigrotzki1994} in a sea of singlets (see below). Up to first order in $\delta J= J_{\parallel} - J_{\cross}$, it is given by:
\begin{eqnarray}
h_{c1} = J_{\perp}-\delta J +\mathcal{O}\left(\frac{{(\delta J)}^2}{J_{\perp}}\right) \nonumber
\end{eqnarray} 

\section{Intensities from DMRG results}
In this section, we provide details of how we have extracted the intensities of the split band and of the main branch of the $t^0$-triplet mode shown in Fig.~4 of the main text. 
The intensity of each mode has been calculated by integrating the DSF between two frequencies: 
 \begin{equation*} I_{split}(M^z) =  \int_{\omega_1}^{\omega_2} S^{zz}_{\pi}(k=0,\omega ) d\omega, 
 \end{equation*} 
\begin{equation*} I_{main}(M^z) = \int_{\omega_2}^{\omega_3} S^{zz}_{\pi}(k=0,\omega ) d\omega. \end{equation*}

For low values of the rung magnetization, the intensity essentially drops to zero between the peak of the split band and the first peak of the main band, and the choice of the three $\omega$ values is obvious. For larger values of the magnetization, the intensity remains finite between the two main peaks, and we have chosen for $\omega_2$ the frequency at the which the minimum is reached (see Fig. \ref{Supplfig4}).

In Fig.~4 of the main text and in Fig.~\ref{Supplfig9} below, the results for different magnetizations have been normalized with respect to the intensity of the main branch at $M^z=0$. In other words, we are plotting $I_{main}(M^z)/I_{main}(M^z=0)$ and $I_{split}(M^z)/I_{main}(M^z=0)$ as a function of the rung magnetization $M^z$.

\begin{figure} 
\centering
\includegraphics[width = 7.5cm ,height  = 7.5cm ,keepaspectratio]{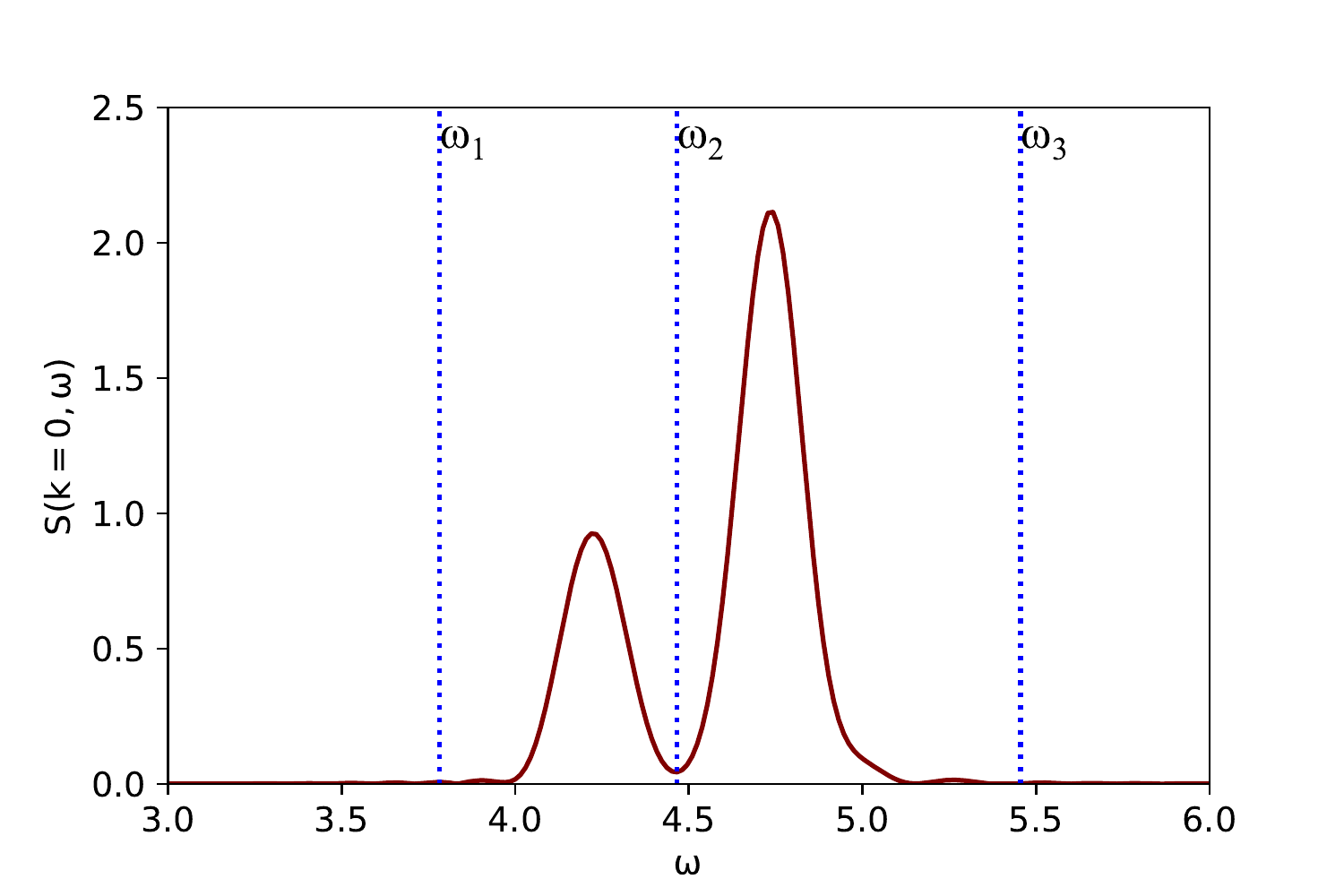}
\caption{Section cut of the anti-symmetric longitudinal component of the DSF for spinladder at k = 0 for rung magnetization $M^z = 0.1$. The dashed lines show the frequencies $\omega_1$, $\omega_2$ and $\omega_3$ chosen to calculate the intensities.}
\label{Supplfig4}
\end{figure}

\section{Dispersion of the intermediate modes}

In this section, we discuss the dispersion of the intermediate modes close to the fully-frustrated case up to second order in $\delta J= J_{\parallel} - J_{\cross}$. To keep the discussion simple, we assume that $J_{\parallel} \ll J_{\perp}$, and we discard all second order terms containing $J_{\perp}$ in the denominator, keeping only those of order $(\delta J)^2/J_{\parallel}$. We have calculated all these terms, and we have checked that, for the parameters relevant for BPCB, their contribution is completely negligible.  

For that purpose, it is convenient to write the Hamiltonian $\cal{H}$ as the sum of the Hamiltonian of the fully-frustrated ladder $H_{FFL}$ and of a perturbation $\delta H$:
\begin{eqnarray}
&&{\cal H} =H_{FFL} +\delta H \nonumber\\
&&H_{FFL} = H_{\perp}+J_{\parallel}\sum_{\text{$\scriptstyle i$}} \left(S_{i,1}+S_{i,2}\right) \cdot \left(S_{i+1,1}+ S_{i+1,2}\right) \nonumber\\
&&\delta H = \sum_{i}\delta H_i \nonumber\\
&& \delta H_i = -\delta J \left(S_{i,1}\cdot S_{i+1,2} + S_{i,2}\cdot S_{i+1,1}\right) \nonumber
\end{eqnarray}
First, we discuss the groundstate of the fully frustrated Hamiltonian $H_{FFL}$. It is given by a product of rung-singlets and is denoted by $\ket{\dots sss \dots}$ where each '$s$' stands for a rung singlet. Its energy is given by: 
\begin{eqnarray}
E^{(0)}_{GS} = -\frac{3J_{\perp}}{4}N_r \nonumber
\end{eqnarray}
where $N_r$ is the number of rungs. This state remains the ground state of $H_{FFL}$ in the presence of a field up to $h_{c1}$, and its energy is of course unaffected by the field.

Upon introducing the perturbation $\delta H$, it is easy to check that there is no first order correction, and no second order correction of order $(\delta J)^2/J_{\parallel}$. The first correction is of order $(\delta J)^2/J_{\perp}$. So $E^{(0)}_{GS}$ will be the reference energy when discussing excitations.

\subsection{Single triplet mode dispersion}

For the fully frustrated ladder, acting with $S^{z}_{j,\pi}$ on the $j^{th}$ rung of the unperturbed ground state (which is given by product of rung singlets), leads to an excited state that we denote by  $\ket{\dots s t^0_{j}s \dots}$. This eigenstate is $N_r$-fold degenerate, and its unperturbed energy is given by
$$E_{t^0}^{(0)}= E^{(0)}_{GS}+J_{\perp}$$
where $ST$ stands for "single triplet".

To estimate the effect of the perturbation $\delta H$ to first order, we need to calculate its matrix elements between two ground states. Since $\delta H_i \ket{\dots s t^0_{i}s \dots}=\ket{\dots s t^0_{i+1}s \dots}$, they read:
\begin{equation*}
\bra{\dots s t^0_{i}s \dots}\delta H \ket{\dots s t^0_{j}s \dots} =  \frac{(\delta J)}{2}\left(\delta_{i,j-1}+\delta_{i,j+1}\right) 
\end{equation*}
The resulting effective hamiltonian can be diagonalized by a Fourier transformation, leading to a momentum dependent first-order contribution given by: 
\begin{eqnarray}
E_{t^0}^{(1)} = \delta J \cos k \nonumber
\end{eqnarray}

Combining contributions of order 0 and 1 results in the excitation energy with respect to the ground state
\begin{eqnarray}
\omega_{t^0}(k) = J_{\perp}+\delta J \cos k +\mathcal{O}\left(\frac{{(\delta J)}^2}{J_{\perp}}\right)\nonumber
\label{0tripletdispersion}
\end{eqnarray}
 
 This dispersion relation is also valid for the $t^{+}$-triplet and $t^{-}$-triplet in the absence of external magnetic field (i.e. $h=0$). When an external magnetic field $h$ is switched on, the dispersion of the $t^{+}$-triplet is shifted down by $h$, and a phase transition occurs when the bottom of this band is equal to zero, leading to the expression $h_{c1}$ quoted earlier.
This occurs at momentum $k=\pi$ for our choice of $J_{\perp}$ and $J_{\parallel}$.

\subsection{Boundstate dispersion}

As discussed in the previous section, upon applying an external magnetic field, there is a transition at $h_{c1}$ from a ground state that is  a product of rung-singlets to a phase where the ground state is a product of singlets with some $t^{+}$-triplets. For obtaining some analytical insight into the DSF plots where we consider low rung magnetization, we work in the vicinity of $h_{c1}$ and for the rest of the discussion assume a groundstate with one $t^{+}$-triplet in a sea of rung-singlets.

From the previous section, we also know that, away from perfect frustration, the minimum of the dispersion for a single $t^{+}$-triplet occurs at $k = \pi$, so that the total momentum of this ground state is equal to $\pi$. Its energy is given by:
\begin{eqnarray}
E_{GS1} &=& E^{(0)}_{GS}+J_{\perp}-\delta J -h\nonumber
\end{eqnarray}
 We retrieve the value for $h_{c1}$ when $E_{GS1} = E^{(0)}_{GS}$. 
\begin{table}[H]
\renewcommand{\arraystretch}{2}
\centering
\caption{Bound states of 2 adjacent triplets characterized by their total spin $S$ and their spin component along $z$ $S^z$}
\begin{tabular}{c|c|c|c}
$S^z$ &$ S = 0 $ & $ S = 1$ & $ S =2$ \\
\hline
\hline
2 & & & $|t^{+}t^{+}\rangle$  \\
1 & & {\large $\frac{|t^{+}t^0\rangle-|t^{0}t^{+}\rangle}{\sqrt{2}}$} & {\large $\frac{|t^{+}t^0\rangle+ |t^0t^{+}\rangle}{\sqrt{2}}$}\\
0 & {\large $\frac{|t^{+}t^{-}\rangle+|t^{-}t^{+}\rangle-|t^0t^0\rangle}{\sqrt{3}}$} & {\large $\frac{|t^{+}t^{-}\rangle-|t^{-}t^{+}\rangle}{\sqrt{2}}$} & {\large $\frac{|t^{+}t^{-}\rangle+|t^{-}t^{+}\rangle+2|t^0t^0\rangle}{\sqrt{6}} $}\\
-1 & & {\large $\frac{|t^{-}t^0\rangle-|t^0t^{-}\rangle}{\sqrt{2}}$} & {\large $\frac{|t^{-}t^0\rangle+|t^0t^{-}\rangle}{\sqrt{2}}$}\\
-2 & & & $|t^{-}t^{-}\rangle$
\end{tabular}
\label{Table2}
\end{table}

When acting with $S^z_{\pi}$ on a site adjacent to an up-triplet, we create a state of the form $|t^{+}t^0\rangle$ or $|t^0t^{+}\rangle$. These states are not eigenstates of $H_{FFL}$ because the neighbouring $t^{+}$- and $t^0$- triplets are coupled by the leg- and cross-couplings of $H_{FFL}$, but they cannot move as long as they are surrounded by singlets \cite{Honecker_Mila_Normand2016}. The eigenstates of the two-triplet problem are listed in Table I. Out of them, only two contain $|t^{+}t^0\rangle$ or $|t^0t^{+}\rangle$. They are given by  $\ket{BS1}\equiv\ket{S = 1, S^z = 1}$ and $\ket{BS2}\equiv\ket{S= 2, S^z = 1}$. Accodingly, we describe the corresponding boundstates of $H_{FFL}$ as $\ket{\psi_{BS1, j}} = \ket{\dots s{(BS1)}_{j,j+1}s \dots}$  and $\ket{\psi_{BS2, j}} = \ket{\dots s{(BS2)}_{j,j+1}s \dots}$. They contain an immobile two-triplet eigenstate on bond $(j,j+1)$ surrounded by rung-singlets. The eigenvalues ($E_{BS1}$ and $E_{BS2}$) corresponding to the boundstates are given as:  
\begin{eqnarray}
E^{(0)}_{BS1} &=& E^{(0)}_{GS} + 2J_{\perp}-J_{\parallel}\nonumber\\
E^{(0)}_{BS2} &=& E^{(0)}_{GS} + 2J_{\perp}+J_{\parallel}\nonumber
\end{eqnarray}

Let us now introduce the perturbation $\delta H$ such that $\delta J \ll J_{\parallel}$. We first discuss its effect on  $\ket{\psi_{BS1, j}}$. The extension to $\ket{\psi_{BS2, j}}$ will be straightforward. 

The first order contribution to the matrix elements of the effective Hamiltonian are given by : 
\begin{eqnarray}
\bra{\psi_{BS1, i}}\delta H\ket{\psi_{BS1,j}} &=& \frac{\delta J}{2} \delta_{i,j}\nonumber
\end{eqnarray} 
So $\delta H$ does not induce any dispersion to first order. This is quite natural since to move a bound state a triplet has to move two rungs apart, and $\delta H$ is just a sum of terms coupling nearest-neighbouring rungs.

To get a dispersion, we have to calculate the second order contributions to the matrix elements of effective Hamiltonian. 
They can be written as $F_{ij} = \sum_{\phi\neq \psi_{BS1}}F_{ij}(\phi)$ where :
\begin{eqnarray}
F_{ij}(\phi) = \frac{\bra{\psi_{BS1 ,i}}\delta H\ket{\phi}\bra{\phi}\delta H\ket{\psi_{BS1,j}}}{E^{(0)}_{BS1}-E^{(0)}_{\phi}} \nonumber
\end{eqnarray}

When acting with one term $\delta H_m$ of $\delta H$ on $\ket{\psi_{BS1,j}}$, the only cases leading to an intermediate state with energy $\mathcal{O}(J_{\parallel})$ with respect to $E^{(0)}_{BS1}$ are:

(i) $m = j-1$. The action of $\delta H_m$ on the boundstate at bond $(j,j+1)$ leads to an intermediate state $\phi$ with a singlet in between a $t^{+}$-triplet and a $t^0$-triplet in a sea of singlets, leading to a contribution to $F_{ij}$ given by:  
\begin{eqnarray}
 -\frac{{(\delta J)}^2}{4J_{\parallel}}\left(\delta_{i,j}+\delta_{i,j-1}\right)\nonumber
\end{eqnarray}

(ii) $m = j+1$. The matrix elements are obtained in a similar way as in case (i) by replacing $\delta_{i,j-1}$ with $\delta_{i,j+1}$, leading to a contribution to $F_{ij}$ given by: 
\begin{eqnarray}
 -\frac{{(\delta J)}^2}{4J_{\parallel}}\left(\delta_{i,j}+\delta_{i,j+1}\right) \nonumber
\end{eqnarray}

The effective Hamiltonian obtained by collecting all terms up to second order can be diagonalized by a Fourier transformation, leading, up to terms of order $\mathcal{O}((\delta J)^2/J_\perp)$, to the following expression for the excitation energy w.r.t. the ground state energy:
\begin{eqnarray}
\omega_{BS1}(k) = J_{\perp}-J_{\parallel}+\frac{3\delta J}{2}-\frac{{(\delta J)}^2}{2J_{\parallel}} \left(1+ \cos k \right)\nonumber
\end{eqnarray}  

Using a similar analysis, the excitation energy corresponding to the bound state of total spin 2 can be shown to be given by: 
\begin{eqnarray}
\omega_{BS2}(k) = J_{\perp}+J_{\parallel}+\frac{\delta J}{2} +\frac{{(\delta J)}^2}{2J_{\parallel}} \left(1+ \cos k\right)\nonumber
\end{eqnarray}

\begin{figure*}
\centering
\includegraphics[width = 18cm , height = 4.5cm , keepaspectratio]{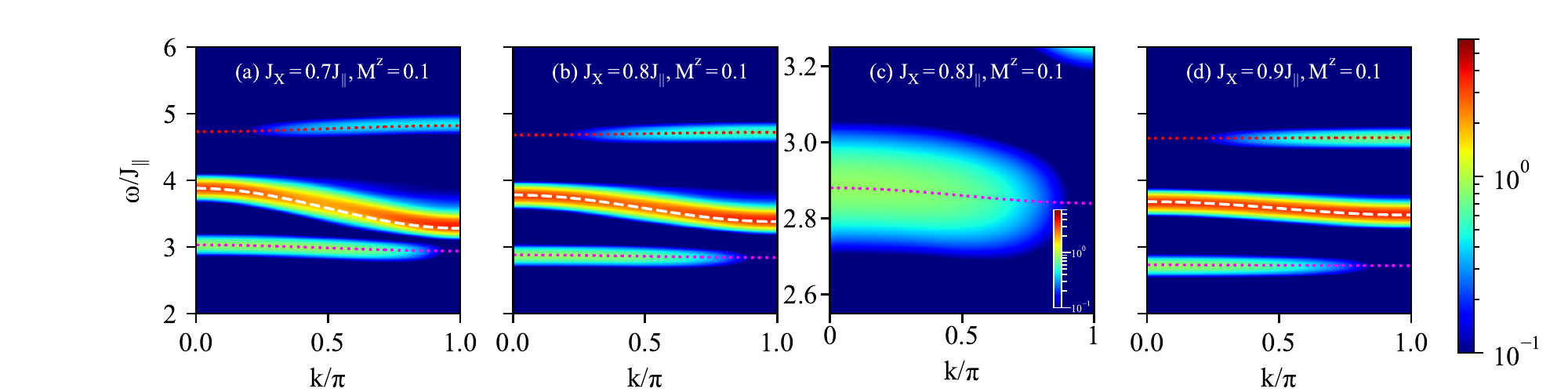}
\caption{Comparison of our analytical expressions for the dispersion relation with the numerical results in the longitudinal antisymmetric DSF that probes the intermediate 0-triplet mode. White dashed line: single-triplet (ST) mode, magenta dashed lined: lower bound state (BS1), red dashed line: upper bound state (BS2). The agreement is perfect close to the the fully frustated case (right panel), and it remains very good further away from full frustration (left panels). The small, second order dispersions of the bound states become visible in the left panels, and they are in quantitative agreement with our second-order calculation. The momentum dependence of the intensity of the bound states is also in agreement with our prediction (see text).}
\label{Supplfig1}
\end{figure*}

\section{Dynamical structure factor of the intermediate-energy modes}

In this section, we build on our perturbative results to calculate the dynamical structure factor of the single triplet excitation and of the bound states. As we shall see, this calculation shows that the momentum of the bound states is shifted by $\pi$, and that the intensity of the bound states is strongly momentum dependent. 

The longitudinal antisymmetric DSF discussed in the main text is given by:
\begin{equation*}
S^{z z}_{\pi}(k,\omega) = \frac{2\pi}{N_r}\sum_{\eta} {\left |\bra{\eta}S^{z}_{\pi}(k)\ket{GS} \right |}^2 \delta(\omega - \omega_{\eta})
\end{equation*}
with $S^{z}_{\pi}(k) = \sum_{j}S^{z}_{\pi,j}e^{ik R_j}$.

To discuss the dynamical structure factor of the single triplet excitation, it is more convenient to consider the ground state just below $h_{c1}$, which is essentially a sea of singlets. The single triplet wave function is just a plane wave built out of $t^0$ triplets, leading to 
\begin{equation*}
S^{z z}_{\pi}(k,\omega) = \frac{2\pi}{N_r} \delta(\omega - \omega_{t^0}(k))
\end{equation*}

By contrast, to discuss the dynamical structure factor of the bound state, since it is necessary to have at least one triplet in the ground state, the convenient starting point is the ground state just above $h_{c1}$, with one $t^{+}$-triplet and momentum $\pi$. Up to corrections of order $(\delta J)^2/J_\perp$, it is given by:
\begin{eqnarray}
\ket{GS1} = \sum_{i}e^{i\pi R_{i}}\ket{\dots s t^{+}_{i}s \dots} \nonumber
\end{eqnarray} 

The action of $S^{z}_{\pi}(k) = \sum_{j}S^{z}_{\pi,j}e^{ik R_j}$ on the ground state $\ket{GS1}$ leads to the state:
\begin{eqnarray}
\ket{\alpha(k)} = S^{z}_{\pi}(k)\ket{GS1} = \sum_{i\neq j}e^{i(\pi R_{i}+kR_{j})}\ket{\dots t^{+}_{i} \dots t^0_{j} \dots}\nonumber
\end{eqnarray}

The bound state of total spin 1 and momentum $q$ is approximately a plane wave built out of local bound states:
\begin{eqnarray}
\ket{\eta_{BS1}(q)} \approx \frac{1}{\sqrt{N_r}} \sum_{l}e^{i q R_{l}}\ket{\dots {(BS1)}_{l,l+1} \dots} \nonumber
\end{eqnarray}
Injecting this state into the Lehmann representation leads to:
\begin{widetext}
\begin{eqnarray}
S^{z z}_{\pi}(k,\omega) & = &\frac{2\pi}{N_r}\sum_{q}{\left |\sum_{i,l}\frac{1}{\sqrt{2N_r}}e^{-iqR_{l}}e^{i(k+\pi)R_{i}}\left(e^{ i k }\delta_{i,l}-e^{-ik}\delta_{i-1,l}\right)\right |}^2 \delta(\omega - \omega_{\eta_{BS1}}(q)) \nonumber\\
&=&\frac{\pi}{N_r} \sum_{q}{|e^{ik}+1|}^2{\left |\delta_{k+\pi,q}\right |}^2 \delta(\omega - \omega_{(BS1)}(q)) \nonumber\\
&=&\frac{2\pi}{N_r}(1+\cos k)\delta(\omega - \omega_{(BS1)}(k+\pi)) \nonumber 
\end{eqnarray}
\end{widetext}

In this case there is a shift by $\pi$ between the momentum of the excitation and that of the DSF. This is a consequence of the $\pi$ momentum of the triplet excitation that condenses at ${h_{c1}}$ because it is involved in the construction of the bound state. By contrast, one can check that there is no shift in the single-triplet excitation, even if it is calculated with respect to $GS1$, because the $t^{+}$ triplet does not combine with this excitation.

A similar analysis can be performed for the bound state of total spin 2 $BS2$, and there is again a shift by $\pi$ of the momentum. The DSF of this excitation is given by:
$$
\frac{2\pi}{N_r}(1-\cos k)\delta(\omega - \omega_{BS2}(k+\pi)) 
$$ 

Interestingly, the amplitude of the DSF for the three branches of excitation has a very different momentum dependence. For the ST excitation, it is momentum independent, for the spin-1 bound state, it vanishes at $k=\pi$, while for the spin-2 bound state, it vanished at $k=0$.

In Fig.\ref{Supplfig1}, we compare our predictions for the intermediate modes  with DMRG results obtained at magnetization $M^z=0.1$ for various levels of frustration close to the fully frustrated ladder. The agreement is overall excellent, including the lack of a first-order dispersion and the progressive development of a small second order dispersion for the bound states, as well as the momentum dependence of their intensity. 

\section{Higher energy modes}

\subsection{Experimental}
\begin{figure}
\centering
\includegraphics[width = 8cm, height = 8cm, keepaspectratio]{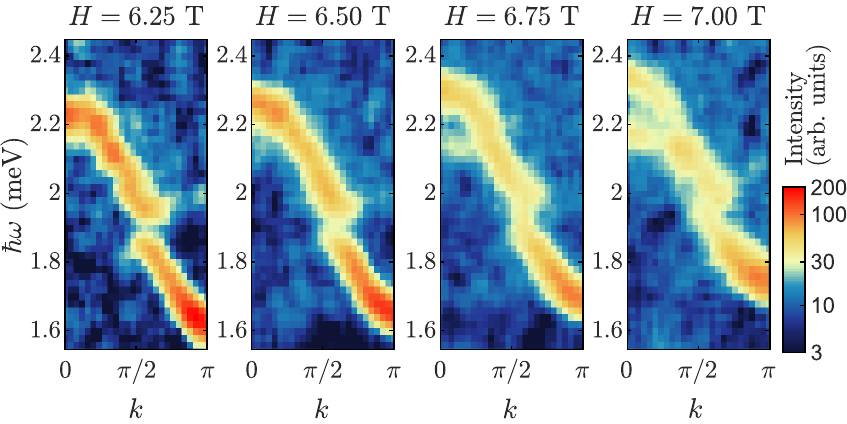}
\caption{Evolution the the INS spectrum measured in BPCB at 0.35 K. In the plotted range of energy transfer only the highest triplet branch of excitations is visible. Upon increasing the magnetic field beyond $H_{c1}$ a distinct splitting is observed at the band maximum. These data were previously published in the supplement to Ref. \cite{Blosser2018}.}
\label{Supplfig5}
\end{figure}

Whilst in the main text only the evolution of the middle triplet band is discussed, a similar branch splitting is observed at the band maximum of the high-energy triplet. Neutron scattering data obtained using $E_\mathrm{i}=4.2$~meV incident energy neutrons, focusing on these high energy excitations, are shown in Fig.~\ref{Supplfig5}. Evidently, upon increasing the magnetic field, the entire band is Zeeman-shifted to higher energies. Further, upon partially magnetizing the ladder beyond $H_{c1}=6.66(6)$~T, a split-off branch at the band maximum gains spectral weight, just as for the middle triplet. Finally, we note that in all data sets obtained using $E_\mathrm{i}=4.2$~meV there is an apparent `line' of low intensity at $\hbar\omega\approx 1.9$~meV, independent of magnetic field or temperature. Most likely, this is an experimental artefact.

\subsection{Theory}
The higher-energy branch is due to the creation of a $t^{-}$-triplet by the transverse anti-symmetric operator $S^{x}_{i,\pi}$. As for the intermediate excitations, the spectrum evolves smoothly between the fully-frustrated case and the non-frustrated case, and the split mode can again be interpreted as a bound state. Let us start by discussing the excitations in the vicinity of the fully frustrated ladder.

\subsubsection{Single $t^{-}$-triplet mode}

The situation is completely similar to the $t^0$-triplet mode, up to the fact that one has to pay the Zeeman energy to create a $t^{-}$-triplet. Accordingly, its dispersion is given by:
\begin{eqnarray}
\omega_{t^{-}}(k) = J_{\perp}+h+\delta J \cos k +\mathcal{O}\left(\frac{{(\delta J)}^2}{J_{\perp}}\right) \nonumber
\end{eqnarray}

\subsubsection{Bound states}

As for the intermediate modes, these modes appear just above $h_{c1}$, when there is a non-zero population of $t^{+}$-triplets in the ground state.  When acting with $S^{x}_{i,\pi}$ on the rung-singlet adjacent to a $t^{+}$-triplet, we create two-site states of the form $\ket{t^{+}t^{-}}$ or $\ket{t^{-}t^{+}}$. These two-site states have an overlap with three of the two-triplet eigenstates of the fully frustrated ladder: $\ket{\overline{BS2}} \equiv  \ket{S=2, S^z=0}$, $\ket{\overline{BS1}} \equiv \ket{S=1, S^z = 0}$ and $\ket{\overline{BS0}} \equiv \ket{S = 0, S^z = 0}$ (seeTable \ref{Table2}). 

Let us define the full eigenstates of $H_{FFL}$ as:
\begin{eqnarray}
\ket{\psi_{\overline{BS2},j}} &=& \ket{\dots s{(\overline{BS2})}_{j,j+1}s \dots}\nonumber\\
\ket{\psi_{\overline{BS1},j}} &=& \ket{\dots s{(\overline{BS1})}_{j,j+1}s \dots}\nonumber\\ 
\ket{\psi_{\overline{BS0},j}} &=& \ket{\dots s{(\overline{BS0})}_{j,j+1}s \dots}\nonumber
\end{eqnarray}
Their energies are given by : 
\begin{eqnarray}
E_{\overline{BS2}}&=& E^{(0)}_{GS}+2J_{\perp}+J_{\parallel}\nonumber \\
E_{\overline{BS1}}&=& E^{(0)}_{GS}+2J_{\perp}-J_{\parallel}\nonumber \\
E_{\overline{BS0}}&=& E^{(0)}_{GS}+2J_{\perp}-2J_{\parallel}\nonumber 
\end{eqnarray}

Using similar arguments as in the discussion of bound states at intermediate energy, one can easily determine the dispersion of these modes including second-order contributions of order in $(\delta J)^2/J_{\parallel}$.  As before, there is no dispersion to first order, and the dispersion to second order is due to intermediate states of energy $\mathcal{O}(J_{\parallel})$ with one singlet between two triplets. The resulting dispersions, valid up to order $\mathcal{O}\left(\frac{{(\delta J)}^2}{J_{\perp}}\right)$, are given by:
\begin{eqnarray}
\omega_{\overline{BS2}}(k) = J_{\perp}+J_{\parallel}+h+\frac{(\delta J)}{2}+\frac{{(\delta J)}^2}{2J_{\parallel}} \left(1+ \cos k\right) \nonumber
\end{eqnarray} 
\begin{eqnarray}
\omega_{\overline{BS1}}(k) =  J_{\perp}+h-J_{\parallel}+\frac{3(\delta J)}{2}-\frac{{(\delta J)}^2}{2J_{\parallel}} \left(1+ \cos k\right)\nonumber
\end{eqnarray}
\begin{eqnarray}
\omega_{\overline{BS0}}(k) &=& J_{\perp}+h-2J_{\parallel}+2\delta J -\frac{{(\delta J)}^2}{4J_{\parallel}} \left(1+ \cos k\right) \nonumber
\end{eqnarray}

\subsubsection{Dynamical Structure Factor of the Higher energy modes}
The calculation is very similar to that of the longitudinal case, and we just quote the results for all modes contributing to the DSF at high energy:
\begin{itemize} 
\item Single $t^{-}$-triplet mode
$$
S^{xx}_{\pi}(k,\omega) = \frac{2\pi}{N_r} \delta(\omega - \omega_{t^{-}}(k))
$$

\item $S=2$ bound state
$$
S^{xx}_{\pi}(k,\omega) = \frac{2\pi}{3N_r}(1-\cos k)\delta(\omega - \omega_{\overline{BS2}}(k+\pi)) 
$$
\item $S=1$ bound state
$$
S^{xx}_{\pi}(k,\omega) = \frac{2\pi}{N_r}(1+\cos k)\delta(\omega - \omega_{\overline{BS1}}(k+\pi)) 
$$
\item $S=0$ bound state
$$
S^{xx}_{\pi}(k,\omega) = \frac{4\pi}{3N_r}(1-\cos k)\delta(\omega - \omega_{\overline{BS0}}(k+\pi)) 
$$
\end{itemize}

The bound state dispersions are again shifted by $\pi$ for the same reason as for the intermediate modes.

\begin{figure*}
\centering
\includegraphics[width = 18cm , height = 4.5cm , keepaspectratio]{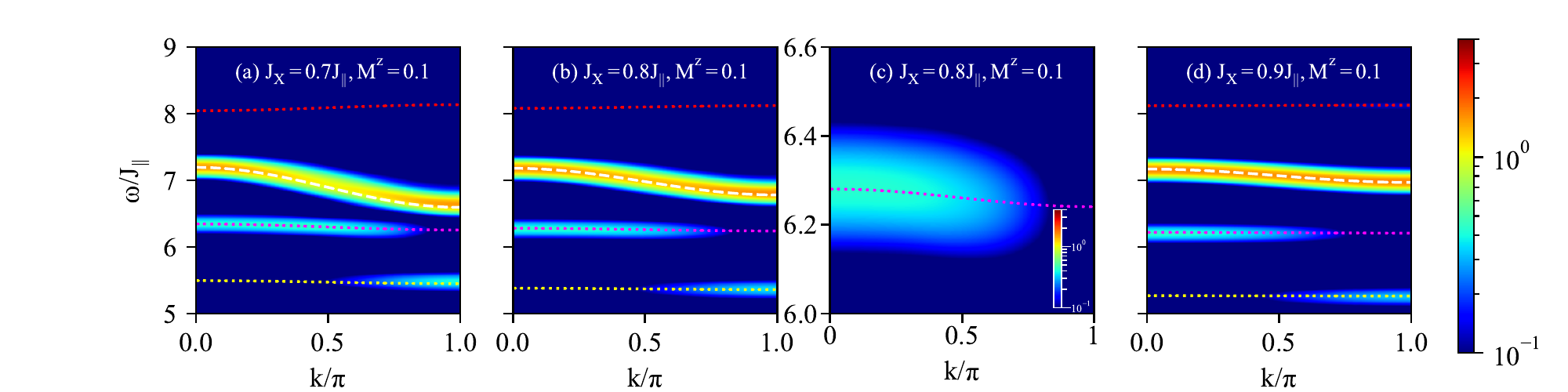}
\caption{Comparison of the analytical expressions for the dispersion relations with the numerical results for the transverse antisymmetric DSF which corresponds to $t^{-}$-triplet mode for rung magnetization $M^z =0.1$. White dashed line: main single $t^{-}$-triplet mode. Then, from top to bottom: Red dashed line: $S=2$ bound state; Magenta dashed line: $S=1$ boundstate; Yellow dashed line: $S=0$ boundstate. The $S=2$ bound state does not appear in the the DMRG results because its weight is too small to be visible. Panel (c) is the magnified version of the bound state with total spin S=1 of panel (b). On this scale, the dispersion acquired by this becomes visible, and the agreement between our analytical expressions and DMRG is very good.}
\label{Supplfig6}
\includegraphics[width = 18cm , height = 4.5cm , keepaspectratio]{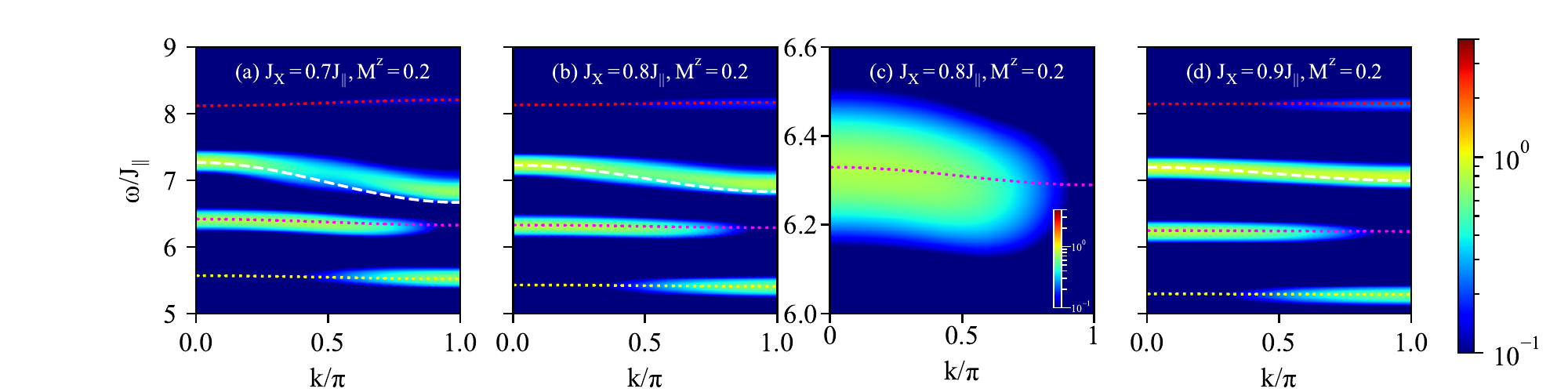}
\caption{Same as Fig. \ref{Supplfig6} but for rung magnetization $M^z =0.2$. The agreement is slightly less good, as it should since the calculation is valid in the limit of vanishing magnetization, but the upper bound state with S=2 becomes visible.}
\label{Supplfig7}
\end{figure*}

We plot the dispersion relations of the bound states ($\omega_{\overline{BS2}}$, $\omega_{\overline{BS1}}$ and $\omega_{\overline{BS0}}$) with a shift of $\pi$ in the transverse anti-symmetric component of DSF (Figs. \ref{Supplfig6} and \ref{Supplfig7}). They agree nicely with the numerical simulations for small $\delta J$. When comparing between the DSF plots with rung magnetization $M^z = 0.1$ and $M^z = 0.2$, we find that the agreement is better with lower magnetization, which is expected since our analytical calculations are strictly valid with one triplet only, when the system is in the vicinity of the critical magnetic field. Note that the upper branch corresponding to the S=2 bound state is not visible in Fig.~\ref{Supplfig6}. This is just due to the small magnetization ($M^z = 0.1$). Indeed, this mode gains weight and becomes visible for a  rung magnetization of $M^z = 0.2$, as can be seen in Fig.~\ref{Supplfig7}.  

\begin{figure*}
\centering
\includegraphics[width = 17.5cm , height = 17.5cm, keepaspectratio]{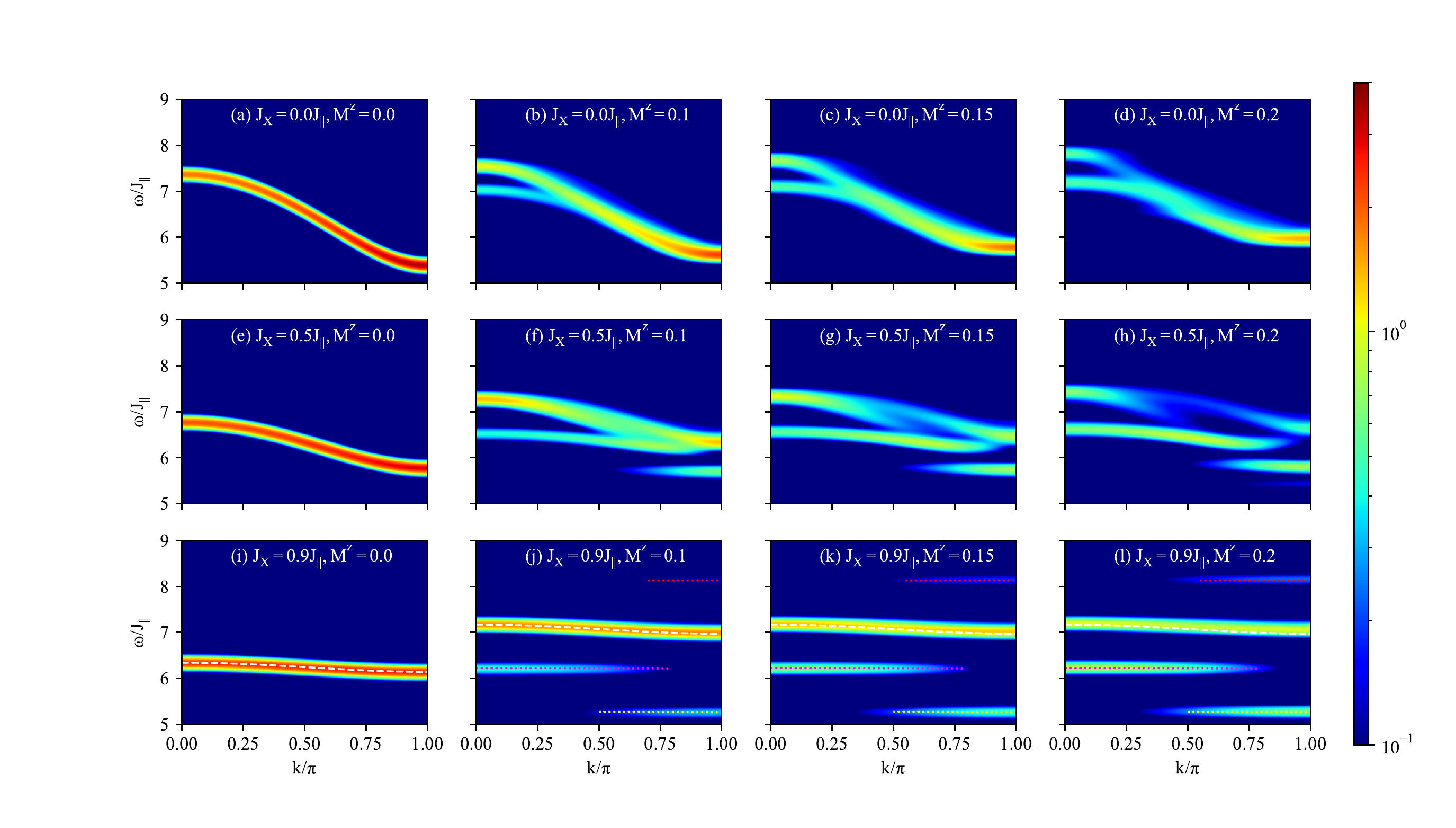}
\caption{Evolution of the transverse antisymmetric DSF($S^{xx}_{\pi}(k,\omega)$) with frustration and magnetization. At strong frustration, the four modes can unambiguously be identified as a dominant single triplet mode and three bound states, as shown by the excellent agreement with the perturbative results close to the fully frustrated case (same convention as in Fig.\ref{Supplfig7}). The spin-1 bound state evolves smoothly into the split mode at k=0 upon suppressing frustration, while the other bound states loose spectral to finally disappear from the spectral function. Unlike in the longitudinal case, the overall position of the triplet bands evolves with the applied magnetic field. We compare the intensity of the two main branches of the unfrustrated case at $k=0$ in Fig.\ref{Supplfig9}.}
\label{Supplfig8}
\end{figure*}

The salient features of the transverse anti-symmetric component of the DSF (see Fig.~\ref{Supplfig8}) can be explained by the dispersion relations we obtained. The $t^{-}$-triplet acquires a dispersion in first order while the bound states are dispersionless up to first order (see lower panels in Fig.~\ref{Supplfig8}). However, the bound states acquire some dispersion as we increase $\delta J$, with a smaller dispersion for the $S=0$ bound state, in agreement with the expressions above.
 As in the case of the analysis of the $t^0$-triplet mode in the main text, we find that the bound states of the frustrated spin ladder smoothly evolve into the high energy modes of the spin ladder. In this case, as we decrease frustration or increase $\delta J$, we find that the bound state with $S=1$ gains significant spectral weight whereas the other bound states lose their spectral weight. Unlike the $t^0$-triplet mode dispersion relation which remained unaffected by the magnetic field, we find that the dispersion relation varies with $h$ in this case because the energy of the $t^{-}$-triplet has a Zeeman contribution.  However, the energy difference between the split mode and the main branch remains unaffected by the magnetic field, as is evident from the dispersion relations.  

\begin{figure}
\centering
\includegraphics[width = 8.5cm, height = 8.5cm, keepaspectratio]{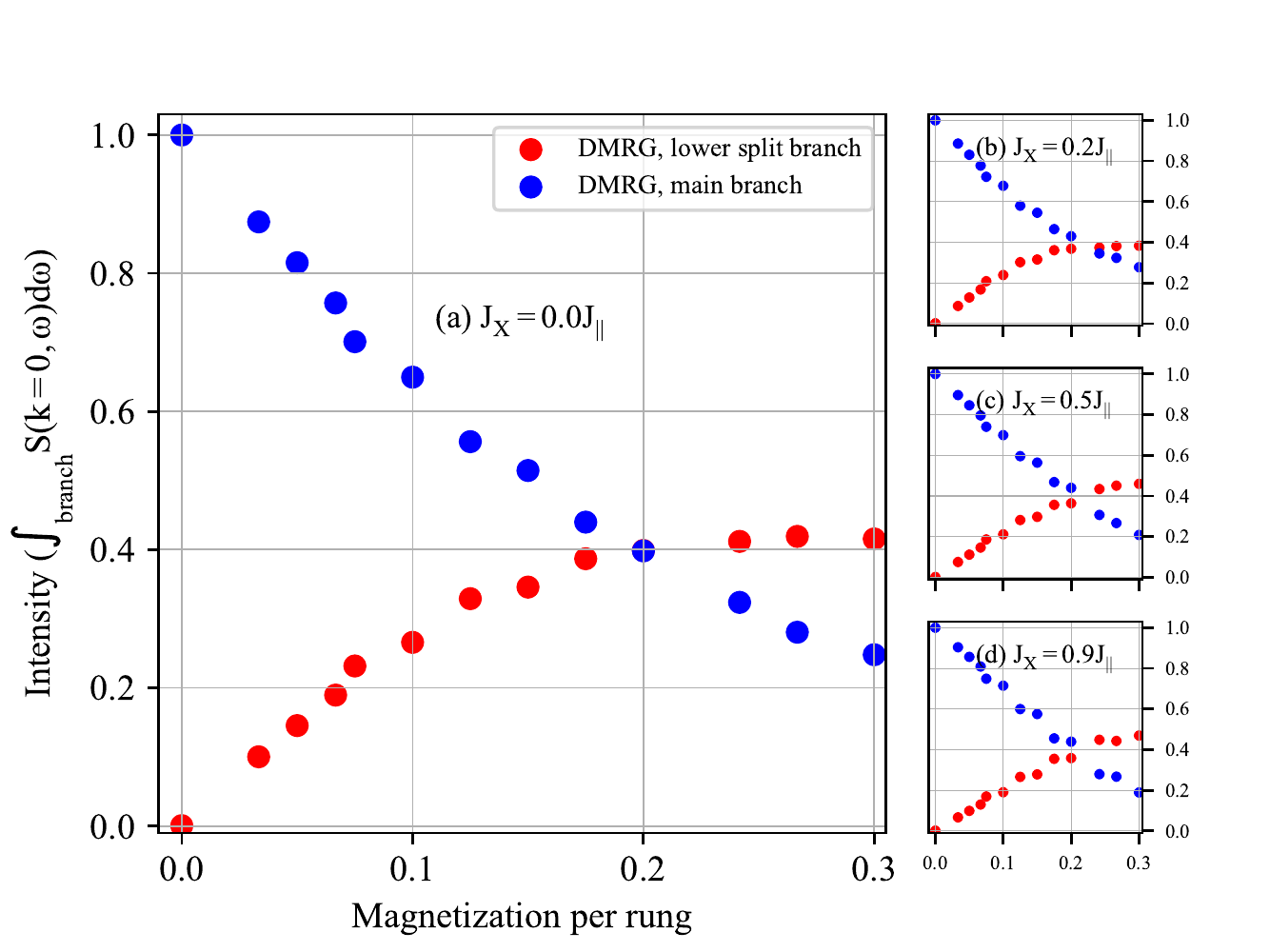}
\caption{Intensity vs. rung magnetization for the $t^{-}$ triplet mode and the lower split mode in the transverse anti-symmetric component of the DSF at $k =0$. At low magnetization, the intensity is linear with very slopes irrespective of the frustration. }
\label{Supplfig9}
\end{figure}

As discussed previously, we followed the same procedure to prepare a plot comparing the intensities of the modes in the vicinity of $k=0$ for the transverse anti-symmetric component of DSF. We find that the modes related to bound states with $S = 2$ and $S=0$ have spectral weights concentrated at $k = \pi$ and have zero spectral weights at $k=0$. Therefore, we considered the intensities (at $k=0$) of the main branch corresponding to the $t^{-}$-triplet excitation and the split excitation corresponding to the bound state with $S=1$ (Fig.~\ref{Supplfig9}). We find that, similar to the case of the $t^{0}$-triplet mode, the intensity of the spectral function corresponding to the bound state increases when the rung magnetization increases, and at the same time the intensity of the single triplet mode decreases. 

\section{Comparison with $t-J$ model}

In this section, we give details about the possible connection between the $t-J$ model on which the ladder has been mapped \cite{Bouillot2011} and our explanation in terms of a bound state of triplets. 

The dynamical structure factor of the spin-$\frac{1}{2}$ ladder has been discussed in  Ref. \cite{Bouillot2011} by mapping the spin ladder to a $t-J$ model. In the intermediate gapless phase, the ground state of the spin ladder in a magnetic field is composed of rung-singlets $\ket{s}$ and rung-triplets $\ket{t^{+}}$. One can map $\ket{s}$  to a pseudo spin-up $\ket{\tilde{\uparrow}}$ and $\ket{t^{+}}$ to a pseudo spin-down $\ket{\tilde{\downarrow}}$, leading to a description of the low-energy dynamics of the the spin ladder in terms of an effective XXZ model. The dynamics of the intermediate energy band corresponds to a $t^{0}$ triplet as pointed out in our discussion before since acting with $S^{z}_{\pi}$ on  $\ket{s}$ leads to a $t^{0}$ triplet in the ladder. If one adopts a fermionic language, the $t^0$ triplet can be considered as a hole, and a $t-J$ model can be formulated to describe the dynamics of these pseudo up-spins, pseudo down-spins and holes:
$$
H_{t-J} = H_{XXZ} + H_t + H_{s-h} + \frac{J_{\perp}+h^{z}}{2}
$$ 
with
\begin{eqnarray}
H_{t} &=& \frac{J_{\parallel}}{2}\sum_{i,\sigma = \tilde{\uparrow},\tilde{\downarrow}} \left(c^{\dagger}_{i,\sigma}c_{i+1,\sigma}+\text{h.c.}\right)\nonumber\\
H_{s-h} &=&- \frac{J_{\parallel}}{2}\sum_{i} n_{i,h}n_{i+1,\tilde{\uparrow}} + n_{i,\tilde{\uparrow}}n_{i+1,h} \nonumber
\end{eqnarray}
where $n$ stands for the number of holes, pseudo-up spins or pseudo down spins depending on the subscript.\\

Interestingly enough from the point of view of bound states, there is an interaction term $H_{s-h}$ between holes and up spins that is {\it attractive}. In the ladder language, this corresponds to the interaction between a $t^{0}$ triplet and a $t^{+}$ triplet.  One of the effects of this term has already been discussed by Bouillot et al \cite{Bouillot2011}. It leads to an asymmetry between the spectra with magnetization $m$ and magnetization $1-m$. To the best of our knowledge, the possibility of the stabilization of a bound state by this interaction has not been discussed previously, but in view of our results, and since this interaction is attractive, we supect that it is this term that is responsible for the formation of a bound state in the context of the $t-J$ model. 

\bibliography{Supplementary_material4.bbl}